\documentclass[11pt]{article}
\usepackage{mathrsfs}
\usepackage{graphicx}

\begin{document}

\title{Metal-Mott Insulator Transition and Spin Exchange
of Two-Component Fermi Gas with Spin-Orbit Coupling in Two-Dimension
Square Optical Lattices}

\author{Beibing Huang\thanks{Corresponding author.
Electronic address: hbb4236@mail.ustc.edu.cn}\\
Department of Experiment Teaching, Yancheng Institute of Technology,
Yancheng, 224051, China
\\ and Shaolong Wan\\
Institute for Theoretical Physics and Department of Modern Physics, \\
University of Science and Technology of China, Hefei, 230026, China}

\maketitle
\begin{center}
\begin{minipage}{120mm}
\vskip 0.8in
\begin{center}{\bf Abstract} \end{center}

{Effects of spin-orbit coupling (SOC) on metal-Mott insulator
transition (MMIT) and spin exchange physics (SEP) of two-component
Fermi gases in two-dimension half-filling square optical lattices
are investigated. In the frame of Kotliar and Ruckenstein slave
boson and the second order perturbation theory, the phase boundary
of paramagnetic MMIT and spin exchange Hamiltonian are calculated.
In addition by adopting two mean-field ansatzs including
antiferromagnetic, ferromagnetic and spiral phases, we find that SOC
can drive a quantum phase transition from antiferromagnet to spiral
phase.}

\end{minipage}
\end{center}

\vskip 1cm

\textbf{PACS} number(s): 03.75.Ss, 51.60.+a, 05.70.Fh

\section{Introduction}

In a crystalline solid spin-orbit coupling (SOC), which occurs
naturally in systems with broken inversion symmetry and makes the
spin degree of freedom respond to its orbital motion, is responsible
for many interesting phenomena, such as magnetoelectric effect
\cite{ide, kato, ganichev}, visionary Datta-Das spin transistor
\cite{datta, jsch}, topological insulator \cite{hasan, kane} and
superconductivity \cite{ludwig, sarma}. Taking topological
superconductivity for example, it has been predicted to occur in
superconductors with a sizable spin-orbit coupling in the presence
of an external magnetic field \cite{sau1, sau2, sau3, sau4, sato}.
In these systems the transition to topological phases requires that
critical magnetic field is much larger than the superconductivity
gap above which an s-wave superconductor is expected to vanish in
the absence of SOC. It is SOC that competes with a strong magnetic
field to give rise to a topological superconducting phase.

As is known to all that ultracold atom systems can be used to
simulate many other systems owing to their many controllable
advantages and operabilities \cite{greiner, moritz, lewenstein}.
Certainly the simulations to SOC, which are generally equivalent to
produce non-abelian gauge potential with optical \cite{op1, op2,
op3} or radio-frequency fields \cite{rf}, are also possible and have
been realized in a neutral atomic Bose-Einstein condensate (BEC) by
dressing two atomic spin states with a pair of lasers
\cite{spileman}. Motivated by such a pioneer experiment and a
practical proposal for generating SOC in ${}^{40}K$ atoms
\cite{k40}, BCS-BEC crossover in the two-component Fermi gases with
SOC have been widely studied \cite{cross3, cross2, cross1, vbshenoy,
melo2, chengang, cross4, melo}.

By contrast in this paper we consider repulsive two-component Fermi
gases with SOC in a two-dimensional square optical lattice, and are
interested in the effects of SOC on metal-Mott insulator transition
(MMIT) and spin exchange physics (SEP) at half filling. Essentially
MMIT of two-component Fermi gases without SOC in an optical lattice
has been realized experimentally \cite{moritz} and is driven by the
competition between hopping term and on-site interaction in the
frame of one-band Hubbard model. When the hopping term dominates,
atoms conduct freely in a lattice and the system is a metal.
Gradually adjusting on-site interaction to an extent that the gain
of the kinetic energy cannot offset the increase of potential
energy, on-site interaction forbids the hopping of atoms and the
system evolves into Mott insulator (MI). In a MI, SEP is correctly
described by quantum antiferromagnetic Heisenberg model, which is
known from the famous $t-J$ model \cite{anderson} expected to offer
a mechanism for high temperature superconductor. Physically this
effective antiferromagnetic coupling between the nearest-neighbor
spins comes from Pauli exclusion principle and the fact that the
hopping of a particle cannot change its spin. Thus to minimize
kinetic energy the nearest-neighbor spins must be antiparallel. In
the presence of SOC, it has two effects on atom hopping. On the one
hand SOC can make atoms move from one site to another site and
corresponds to an effective hopping term, so it definitely has
important effects on MMIT in view of the above statement. On the
other hand, the effective hopping induced by SOC is spin-flipped to
support the nearest-neighbor spins parallel. From this viewpoint SOC
also dramatically changes SEP. Furthermore it is likely that when
the strength of SOC is beyond certain critical value, the system
will show a quantum phase transition from antiferromagnetic to other
magnetic states.

This paper is organized as follows. In section 2, we firstly derive
Hubbard Hamiltonian in two-dimensional square optical lattices with
SOC, then by using Kotliar-Ruckenstein (KR) slave bosons
\cite{kotliar} we investigate paramagnetic MMIT, i.e. Brinkman-Rice
phase transition \cite{rice}. In section 3 under the limit of large
on-site repulsion and using the second order perturbation theory
spin exchange Hamiltonian is obtained. By making mean field
approximations we find that the ground state of the system is either
antiferromagnetic or spiral depending on the relative magnitude of
hopping term and strength of SOC and a quantum phase transition
happens between them. The conclusions are given in section 4.

\section{Metal-MI Phase Transition with SOC}

The Hamiltonian of the system we consider is
\begin{eqnarray}
H=\int d^2\vec{r}\left\{\sum_{\alpha, \beta}
\Psi^{\dag}_{\alpha}(\vec{r})\left[\frac{\vec{p}^2}{2m} +
V_{OL}(\vec{r})+ \lambda (\sigma_x p_y - \sigma_y p_x)
\right]\Psi_{\beta}(\vec{r})+g\Psi^{\dag}_{\uparrow}(\vec{r})
\Psi^{\dag}_{\downarrow}(\vec{r}) \Psi_{\downarrow}(\vec{r})
\Psi{\uparrow}(\vec{r})\right\},  \label{1}
\end{eqnarray}
where a Fermi atom of mass $m$ for spin $\alpha$ is described by the
field operators $\Psi_{\alpha}(\vec{r})$ and $V_{OL}(\vec{r})$ is
optical potential for two-dimensional square lattices. $\lambda$,
$g(>0)$ and $\vec{\sigma}$ represent the strength of Rashba SOC,
two-body contact interaction and Pauli matrix respectively. When
temperature is very low and filling factor is not too high, all
atoms are constrained into the lowest band of the optical lattice.
Expanding the field operator in terms of the Wannier functions
$\Psi_{\alpha}(\vec{r})=\sum_i a_{i\alpha}w(\vec{r}-\vec{R_i})$,
where $a_{i\alpha}$ is the annihilation operator for an atom of spin
$\alpha$ in site $\vec{R_i}$, and only retaining on-site interaction
and nearest neighbor hopping, we find
\begin{eqnarray}
H=\sum_{<i,j>}a_{i\alpha}^{\dag}t_{ij}^{\alpha\beta}a_{j\beta}
+U\sum_ia_{i\uparrow}^{\dag}a_{i\downarrow}^{\dag}a_{i\downarrow}a_{i\uparrow},
\label{2}
\end{eqnarray}
where the hopping term $t_{ij}^{\alpha\beta}$ is a $2\times2$ matrix
and its elements are
$t_{ij}^{\uparrow\uparrow}=t_{ij}^{\downarrow\downarrow}=-t$,
$t_{ij}^{\uparrow\downarrow}=-\left[t_{ij}^{\downarrow\uparrow}\right]^{\ast}=
\Gamma_{i,j}^x-i\Gamma_{i,j}^y$. These parameters $t$,
$\Gamma_{i,j}^x$, $\Gamma_{i,j}^y$ and $U$ are related to the
Wannier function as follows
\begin{eqnarray}
t&=&-\int d^2\vec{r}w(\vec{r}-\vec{R_i})\left[\frac{\vec{p}^2}{2m} +
V_{OL}(\vec{r})\right]w(\vec{r}-\vec{R_j}), \nonumber\\
U&=&g\int d^2\vec{r}w(\vec{r}-\vec{R_i})w(\vec{r}-\vec{R_i})
w(\vec{r}-\vec{R_i})w(\vec{r}-\vec{R_i}),\nonumber\\
\Gamma_{i,j}^x&=&\lambda\int d^2\vec{r}w(\vec{r}-\vec{R_i})
\frac{\partial}{\partial x}w(\vec{r}-\vec{R_j}),\nonumber\\
\Gamma_{i,j}^y&=&\lambda\int d^2\vec{r}w(\vec{r}-\vec{R_i})
\frac{\partial}{\partial y}w(\vec{r}-\vec{R_j}).\label{3}
\end{eqnarray}
From above expressions (\ref{3}) and the symmetry of Wannier
function, $\Gamma_{i,j}^x$, $\Gamma_{i,j}^y$ satisfy the relations
$\Gamma_{i,j}^x=\Gamma_{i,j}^y=-\Gamma_{j,i}^x =-\Gamma_{j,i}^y$.
Moreover $\Gamma_{i,j}^x=0$ if $i, j$ are nearest neighbor along $y$
direction and $\Gamma_{i,j}^y=0$ if $i, j$ are nearest neighbor
along $x$ direction. For convenience the parameter $\Gamma$ is
defined $\Gamma=|\Gamma_{i,j}^x|=|\Gamma_{i,j}^y|$ to represent the
strength of SOC.

It is well known that MMIT is a phenomenon of strong correlation. In
terms of strong correlation, apart from some numerical methods, such
as dynamical mean-field theory \cite{kot}, a few analytical methods
are also available. The first is Gutzwiller variational wave
function \cite{gutzwiller}. In this method to make the calculation
tractable, one has to introduce the Gutzwiller approximation which
is basically at the mean-field level. Although this method is
successful to predict the existence of MMIT, it still has some
disadvantages from variational and mean-field approximations.
Another method is KR slave bosons \cite{kotliar}. It exactly
reproduces the results of Gutzwiller approximation at the
saddle-point level and can be improved systematically by considering
fluctuations around the saddle point \cite{langvin}. Hence below we
adopt slave bosons to study MMIT with SOC, although our results is
at the saddle-point level.

For two-component Fermi gases, the Hilbert space for every lattice
site $i$ consists of four states $|0>_i$,
$|\alpha>_i=a_{i\alpha}^{\dag}|0>_i$ and $|\uparrow,
\downarrow>_i=a_{i\uparrow}^{\dag}a_{i\downarrow}^{\dag}|0>_i$. In
the representation of KR slave bosons, in addition to original
fermions, a set of four bosons $e$, $d$, $p_{\alpha}$ for every
lattice site are introduced so that $|0>_i=e_i^{\dag}|vac>$,
$|\alpha>_i=p_{i\alpha}^{\dag}a_{i\alpha}^{\dag}|vac>_i$ and
$|\uparrow,
\downarrow>_i=d_i^{\dag}a_{i\uparrow}^{\dag}a_{i\downarrow}^{\dag}|vac>_i$,
where $|vac>$ is the vacuum state after introducing slave bosons. It
is easily found that $e_i^{\dag}e_i$, $d_i^{\dag}d_i$ and
$p_{i\alpha}^{\dag}p_{i\alpha}$ represent the projectors on the
empty, doubly occupied and singly occupied site. Due to the fact
that the introduction of bosons enlarges the Hilbert space of every
site to contain some unphysical states, such as
$e_{i}^{\dag}e_{i}^{\dag}|vac>_i$ etc., we must impose three
constraints
$e_i^{\dag}e_i+p_{i\alpha}^{\dag}p_{i\alpha}+d_i^{\dag}d_i=1$,
$a_{i\alpha}^{\dag}a_{i\alpha}=p_{i\alpha}^{\dag}p_{i\alpha}+d_i^{\dag}d_i$.
In terms of these bosons and considering above constraints the
Hamiltonian (\ref{2}) is reformulated into
\begin{eqnarray}
H=\sum_{<i,j>}a_{i\alpha}^{\dag}z_{i\alpha}^{\dag}t_{ij}^{\alpha\beta}z_{j\beta}a_{j\beta}
+U\sum_id_{i}^{\dag}d_{i}, \label{4}
\end{eqnarray}
with
\begin{eqnarray}
&&z_{i\alpha}=(1-d_i^{\dag}d_i-p_{i\alpha}^{\dag}p_{i\alpha})^{-1/2}
\overline{z}_{i\alpha}
(1-e_i^{\dag}e_i-p_{i-\alpha}^{\dag}p_{i-\alpha})^{-1/2},\nonumber\\
&&\overline{z}_{i\alpha}=e^{\dag}_ip_{i\alpha}+p_{i-\alpha}^{\dag}d_i.
\label{5}
\end{eqnarray}
As claimed by KR, the substitution $z_{i\alpha}$ for
$\overline{z}_{i\alpha}$ ensures $z_{i\alpha}^{\dag}z_{j\beta}=1$ to
recover the results in the limit $U=0$ at the saddle-point
approximation.

The partition function $Z$ can be written as a functional integral
over the fermion and boson operators
\begin{eqnarray}
Z=\int\mathscr{D}a_{\alpha}\mathscr{D}e\mathscr{D}p_{\alpha}\mathscr{D}d
\prod_{i\sigma}d\lambda_i^{(1)}d\lambda_{i\alpha}^{(2)}\exp{[-\int_0^{\beta}\mathcal
{L}(\tau)d\tau]},\label{6}
\end{eqnarray}
where the Lagrangian $\mathcal {L}(\tau)$ is
\begin{eqnarray}
\mathcal {L}(\tau)&=&\sum_ie_i^{\dag}[\frac{\partial}{\partial
\tau}+\lambda_i^{(1)}]e_i + d_i^{\dag}[\frac{\partial}{\partial
\tau}+U+\lambda_i^{(1)}-\lambda_{i\uparrow}^{(2)}-\lambda_{i\downarrow}^{(2)}]d_i
+p_{i\alpha}^{\dag}[\frac{\partial}{\partial
\tau}+\lambda_i^{(1)}-\lambda_{i\alpha}^{(2)}]p_{i\alpha}\nonumber\\&&
+ \sum_{<i,j>}a_{i\alpha}^{\dag}\left[ (\frac{\partial}{\partial
\tau}+\lambda_{i\alpha}^{(2)}-\mu)\delta_{\alpha\beta}\delta_{ij}+z_{i\alpha}^{\dag}
t_{ij}^{\alpha\beta}z_{j\beta}\right]a_{j\beta}-\lambda_i^{(1)},
\label{7}
\end{eqnarray}
and $\mu$, $\lambda_i^{(1)}$, $\lambda_{i\alpha}^{(2)}$ are the
chemical potential and Lagrange multipliers, respectively.

Assuming uniform and static boson operators and Lagrange
multipliers, i.e. at the saddle point, one can integrate over
fermion operators and obtain for thermodynamic potential of single
site
\begin{eqnarray}
\Omega=\lambda^{(1)}(e^2+d^2+p_{\alpha}^2-1)+Ud^2-\lambda_{\alpha}^{(2)}
(d^2+p_{\alpha}^2)-\frac{1}{\beta N}\sum_{k\alpha}\ln{[1+e^{-\beta
E_{k\alpha}}]} \label{8}
\end{eqnarray}
with $E_{k\alpha}=\left[\epsilon_{k\uparrow}+\epsilon_{k\downarrow}+
\alpha\sqrt{(\epsilon_{k\uparrow}-\epsilon_{k\downarrow})^2+4z_{\uparrow}^2z_{\downarrow}^2\Gamma_k^2}\right]/2
$,
$\epsilon_{k\alpha}=\epsilon_kz_{\alpha}^2-\mu+\lambda_{\alpha}^{(2)}$,
$\epsilon_k=-2t(\cos{k_xa} + \cos{k_ya})$,
$\Gamma_k=2\Gamma\sqrt{\sin^2{k_xa}+\sin^2{k_ya}}$. It is to be
noted that $a$, $N$ are lattice length and the number of lattice
site, and wavevector $k$ belongs to two dimensional Brillouin zone.
At this time the seven parameters $e$, $p_{\alpha}$, $d$,
$\lambda^{(1)}$ and $\lambda^{(2)}_{\alpha}$ are obtained by
minimizing $\Omega$, and the chemical potential at half filling by
thermodynamic relation $-\frac{\partial \Omega}{\partial \mu}=1$.
These equations are called saddle-point and number equations.

From $\frac{\partial \Omega}{\partial \lambda^{(1)}} =
\frac{\partial \Omega}{\partial \lambda^{(2)}_{\alpha}}=0$ and
$-\frac{\partial \Omega}{\partial \mu}=1$, one can get $e^2=d^2$.
Supposing paramagnetic solution $p_{\uparrow}^2=p_{\downarrow}^2$,
then $p^2=\frac{1}{2}-d^2$,
$z_{\uparrow}^2=z_{\downarrow}^2=z^2=8d^2(1-2d^2)$. According to
$\frac{\partial \Omega}{\partial p_{\alpha}} =\frac{\partial
\Omega}{\partial e}=\frac{\partial \Omega}{\partial d}=0$,
$\lambda_{\uparrow}^{(2)}= \lambda_{\downarrow}^{(2)}=\frac{U}{2}$,
$\epsilon_{k\uparrow}=\epsilon_{k\downarrow}$,
$E_{k\alpha}=\left[\epsilon_{k}+
\alpha\Gamma_k\right]z^2-\mu+\frac{U}{2}$,
$\lambda^{(1)}=\frac{U}{2}-16\xi d^2(3-4d^2)$ with
\begin{eqnarray}
\xi=\frac{1}{N}\sum_k \left[ \frac{\epsilon_k+\Gamma_k}{e^{\beta
E_{k\uparrow}}+1}+\frac{\epsilon_k-\Gamma_k}{e^{\beta
E_{k\downarrow}}+1}\right]. \label{9}
\end{eqnarray}
Substituting above relations into saddle-point and number equations,
one still has two equations satisfied by $\mu$ and $d$
\begin{eqnarray}
&&\frac{1}{N}\sum_k \left[\frac{1}{e^{\beta
E_{k\uparrow}}+1}+\frac{1}{e^{\beta E_{k\downarrow}}+1}
\right]=1,\nonumber\\
&&U+8\xi (1-4d^2)=0.  \label{10}
\end{eqnarray}

At zero temperature, in the frame of KR slave bosons, $d^2=0$
corresponds to the vanishing of the number of doubly occupied sites
and indicates that the system is undergoing a MMIT. From this
criterion one has numerically solved the equations (\ref{10}). The
numerical results suggest the chemical potential is still fixed at
$\mu=U/2$. Hence at zero temperature
\begin{eqnarray}
\xi=\frac{1}{N}\sum_k
\left\{(\epsilon_k+\Gamma_k)\Theta[-(\epsilon_k+\Gamma_k)]+
(\epsilon_k-\Gamma_k)\Theta[-(\epsilon_k-\Gamma_k)]\right\}.
\label{11}
\end{eqnarray}
and the phase boundary of MMIT is
\begin{eqnarray}
U=-8\xi, \label{12}
\end{eqnarray}
where $\Theta(x)$ is Heaviside step function. Without SOC,
$\xi=\frac{2}{N}\sum_k \epsilon_k\Theta[-\epsilon_k]$ and the phase
boundary (\ref{12}) is the same as the result in \cite{kotliar}. In
Fig.1 the phase boundary of MMIT is shown. From Fig.1, very
explicitly SOC stabilizes the MI, which is consistent with the fact
that SOC can be regarded as an effective hopping term. Besides
instead of adjusting $t$ MMIT can also be driven by changing SOC, so
one has found another way to realize the MMIT.

\section{Spin Exchange and Magnetic Phase Transition with SOC}

As demonstrated in the section 2, at half filling when $U>>t$ and
$U>>\Gamma$, the hopping of atoms are forbidden and the system
evolves into MI with spin $S=\frac{1}{2}$ for every lattice site. In
the MI we could regard $t$ and $\Gamma$ as perturbations. In the
limit of $t=\Gamma=0$ the energy of the system does not depend on
the spin orientations on different sites. When $t$, $\Gamma$ are
finite but small, we expect that we still have spin $S=\frac{1}{2}$
in each site, but atom hopping processes induce effective
interactions between these spins, usually called spin exchange
interaction \cite{iim}. To construct an effective spin exchange
Hamiltonian for this system, we note that in the second order in
$t$, $\Gamma$ it can be written as a sum of interaction terms for
all nearest neighbor sites. These pairwise interactions can be found
by solving a two-site problem in the second order in $t$, $\Gamma$.

The ground state manifold for two-site problem with one atom in each
site composes of four degenerate zero-energy states
\begin{eqnarray}
&&|1>=|\uparrow>_i|\uparrow>_j, |2>=|\uparrow>_i|
\downarrow>_j,\nonumber\\
&&|3>=|\downarrow>_i|\uparrow>_j, |4>=|\downarrow>_i|\downarrow>_j,
\label{3a}
\end{eqnarray}
with $i, j$ labelling two sites. The first order perturbation theory
takes us out of the ground state manifold and can be neglected. In
the second order atom hoppings can connect all four states by two
intermediate states $|5>=|\uparrow, \downarrow>_i|0>_j$ and
$|6>=|0>_i|\uparrow, \downarrow>_j$. To find spin exchange
Hamiltonian we need calculate all matrix elements \cite{iim}
\begin{eqnarray}
M_{ab}=\sum_c\frac{<a|H_k |c><c| H_k|b>}{E^0_b-E^0_c}, \label{3b}
\end{eqnarray}
where states $|a>,|b>$ and $|c>$ respectively belong to ground state
manifold and intermediate states with $E^0$ representing eigenenergy
of corresponding state in the zeroth order. The calculation is very
direct and when $i, j$ are nearest neighbor along $x$ direction, we
have
$M_{11}=M_{14}=M_{41}=M_{44}=\frac{2\Gamma_{ij}^x\Gamma_{ji}^x}{U}$,
$M_{12}=M_{21}=M_{24}=M_{42}=\frac{2t\Gamma_{ji}^x}{U}$,
$M_{13}=M_{31}=M_{34}=M_{43}=\frac{2t\Gamma_{ij}^x}{U}$,
$M_{22}=-M_{23}=-M_{32}=M_{33}=-\frac{2t^2}{U}$. According to
spectral representation of an operator, magnetic Hamiltonian of
two-site problem is $H_{i,j}=\sum_{a,b}|a>M_{ab}<b|$. Making
substitutions $|1>\rightarrow
a^{\dag}_{i\uparrow}a^{\dag}_{j\uparrow}$, $|2>\rightarrow
a^{\dag}_{i\uparrow}a^{\dag}_{j\downarrow}$, $|3>\rightarrow
a^{\dag}_{i\downarrow}a^{\dag}_{j\uparrow}$, $|4>\rightarrow
a^{\dag}_{i\downarrow}a^{\dag}_{j\downarrow}$ and using algebra of
spin operator $\vec{S}_i=\frac{1}{2}a^{\dag}_{i\alpha}
\vec{\sigma}_{\alpha\beta}a_{i\beta}$, we get
\begin{eqnarray}
H_{i,j}=\frac{4t^2}{U}\vec{S}_i\cdot\vec{S}_j+\frac{8t\Gamma_{i,j}^{x}}{U}
(S_{ix}S_{jz}-S_{iz}S_{jx})
+\frac{4\Gamma_{i,j}^{x}\Gamma_{j,i}^{x}}{U}
(S_{iz}S_{jz}+S_{ix}S_{jx}-S_{iy}S_{jy}).\label{4'}
\end{eqnarray}
By the same procedure, when $i, j$ are nearest neighbor along $y$
direction we get
\begin{eqnarray}
H_{i,j}=\frac{4t^2}{U}\vec{S}_i\cdot\vec{S}_j+\frac{8t\Gamma_{i,j}^{y}}{U}
(S_{iy}S_{jz}-S_{iz}S_{jy})
+\frac{4\Gamma_{i,j}^{y}\Gamma_{j,i}^{y}}{U}
(S_{iz}S_{jz}+S_{iy}S_{jy}-S_{ix}S_{jx}).\label{5'}
\end{eqnarray}
Thus spin exchange Hamiltonian of the whole system is
\begin{eqnarray}
H_{se}=\sum_{<i,j>}H_{i,j}. \label{6'}
\end{eqnarray}

Some comments about (\ref{6'}) is following. If $\Gamma=0$ $H_{se}$
describes isotropic quantum antiferromagnet, consistent with $t-J$
model. When $\Gamma\neq 0$, main effect of SOC is to break spin
conservation by two ways, one of which, corresponding to the second
term in (\ref{4'}) and (\ref{5'}), flips one spin of two nearest
neighbor sites, while the other flips simultaneously two spins
corresponding to the third term in (\ref{4'}) and (\ref{5'}). Thus
the antiferromagnetic state will be unstable when the strength of
SOC $\Gamma$ is beyond certain critical value.

Now we decide the ground state of the system at mean-field level.
This corresponds to regard quantum spin operator $\vec{S}_i$ as a
classical vector. The first mean-field ansatz including
ferromagnetic and antiferromagnetic states is that spin
configurations in two sublattices of a square lattice take different
values specified respectively by coordinate angle $(\vartheta,
\varphi)$ and $(\gamma, \delta)$ with $0\leq\vartheta,\gamma<\pi$
and $0\leq\varphi,\delta<2\pi$, the mean-field energy scaled by $U$
is
\begin{eqnarray}
E=8N^2\left[(\widetilde{t}^2-\widetilde{\Gamma}^2)\cos\vartheta\cos\gamma+
\widetilde{t}^2 \sin\vartheta\sin\gamma\cos(\varphi-\delta)\right],
\label{7'}
\end{eqnarray}
where a $2N\times 2N$ lattice is assumed and $\widetilde{t}=t/U$,
$\widetilde{\Gamma}=\Gamma/U$. Easily found that energy only depends
on the difference of $\varphi$ and $\delta$, for convenience we can
choose $\delta=0$. Owing to factor $\sin\vartheta\sin\gamma\geq 0$,
the minimization of energy leads to $\varphi=\pi$. Minimizing energy
about $\vartheta, \gamma$, we get
\begin{eqnarray}
&&(\widetilde{t}^2-\widetilde{\Gamma}^2)\sin\vartheta\cos\gamma+
\widetilde{t}^2 \cos\vartheta\sin\gamma=0,\nonumber\\
&&(\widetilde{t}^2-\widetilde{\Gamma}^2)\cos\vartheta\sin\gamma+
\widetilde{t}^2 \sin\vartheta\cos\gamma=0. \label{8'}
\end{eqnarray}
Equations (\ref{8'}) have two sets of solution $\vartheta=\gamma=0$
and $\vartheta=\gamma=\pi/2$. The first solution corresponds to
ferromagnet along $z$ direction with
$E_{FE}=8N^2(\widetilde{t}^2-\widetilde{\Gamma}^2)$ and the second
corresponds to antiferromagnet along $x$ direction with
$E_{AF}=-8N^2\widetilde{t}^2$. If $E_{FE}<E_{AF}$ ground state is
ferromagnetic, on the contrary ground state is antiferromagnetic.
Thus this mean-field ansatz predicts a phase transition from
antiferromagnet to ferromagnet and the critical point is
$E_{FE}=E_{AF}$, i.e. $\widetilde{\Gamma} =\sqrt{2}\widetilde{t}$.

The motivation of the second mean-field ansatz comes from $t=0$
limit in spin exchange Hamiltonian (\ref{6'}). Letting $t=0$ the
classical spin configuration minimizing energy satisfies three
conditions: (1) $z$ components of all spins are equal; (2) for a
random chain along $x$ direction $x$ components of all spins are
equal but $y$ component must be alternating; (3) for a random chain
along $y$ direction $y$ components of all spins are equal but $x$
component must be alternating. Such spin configuration, which we
call spiral phase and shown in Fig.2, is permissible in a square
lattice. From above three conditions if coordinate angle $(\theta,
\phi)$ of a spin in the lattice is specified, energy of the system
is
\begin{eqnarray}
E_{SP}=8N^2(\widetilde{t}^2\cos^2\theta-\widetilde{\Gamma}^2),
\label{9'}
\end{eqnarray}
and its minimization gives rise to $\theta=\pi/2$, $E_{SP}=
-8N^2\widetilde{\Gamma}^2$. Comparing $E_{SP}$ with $E_{FE}$ we find
that the ferromagnetic state is always a metastable state. As a
result phase transition predicted by the first mean-field ansatz
does not exist, we get a phase transition from antiferromagnet to
spiral phase with critical point $\widetilde{\Gamma}=\widetilde{t}$.
Fig.1 also shows magnetic phase diagram in terms of such two
mean-field ansatzs. Physically the metastability of ferromagnetic
state is attributed to the fact that SOC breaks spin conservation.

\section{Conclusions}

In conclusion we have discussed MMIT and SEP of two-component Fermi
gases with SOC in two-dimensional half-filling square optical
lattices in the frame of KR slave bosons and second-order
perturbation theory. Comparing with the case without SOC, SOC not
only enlarges the region of MI in the phase diagram and introduces
another way to realize MMIT, but also dramatically affects SEP due
to SOC breaking spin conservation. Importantly by adopting two
mean-field ansatzs we find that SOC can drive a phase transition
from antiferromagnet to spiral phase. Experimentally this phase
transition can be observed by either adjusting optical lattices to
suppress the hopping term or decreasing the strength of SOC.

\section*{Acknowledgement}

The work was supported by National Natural Science Foundation of
China under Grant No. 10675108. The author Huang also thanks
Foundation of Yancheng Institute of Technology under Grant No.
XKR2010007.

\begin{figure}[htbp]
\centering
\includegraphics[width=7.5cm, height=6.0cm]{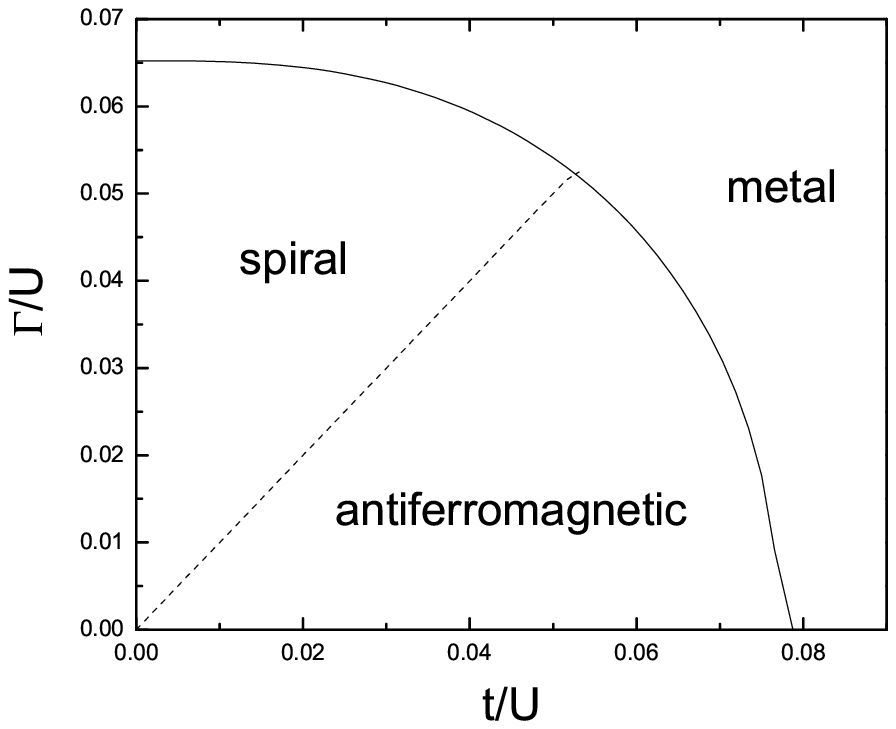}
\caption{The phase diagram of a repulsively interacting
two-component Fermi gas with spin-orbit coupling in a square optical
lattice. The solid line is the phase boundary of metal-Mott
insulator transition, while the dashed line is one of
antiferromagnetic-spiral phases.} \label{fig.2}
\end{figure}

\begin{figure}[htbp]
\centering
\includegraphics[width=7.5cm, height=6.0cm]{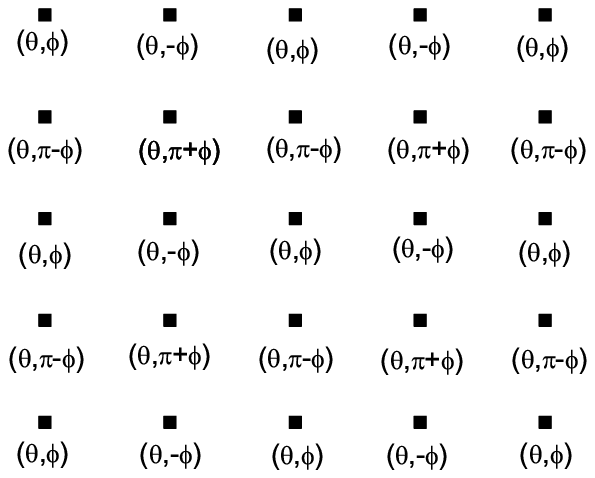}
\caption{Spin configuration of spiral phase is shown with small
squares representing lattice sites. $(\theta, \phi)$, $(\theta,
-\phi)$, $(\theta, \pi-\phi)$ and $(\theta, \pi+\phi)$ are
coordinate angles of spins.} \label{fig.1}
\end{figure}


\begin{thebibliography}{99}
\bibitem{ide} I. E. Dzyaloshinkii, Sov. Phys. JETP 10, 628 (1959).
\bibitem{kato} Y. Kato, R. C. Myers, A. C. Gossard and D. D.
Awschalom, Phys. Rev. Lett. 93, 176601 (2004).
\bibitem{ganichev} S. D. Ganichev, S. N. Danilov, P. Scheider, V. V. Belkov,
L. E. Golub, W. Wegscheider, D. Weiss and W. Prettl, arXiv:0403641.
\bibitem{datta} S. Datta and B. Das, Appl. Phys. Lett. 56, 665 (1990).
\bibitem{jsch} J. Schliemann, J. C. Egues and D. Loss, Phys. Rev. Lett. 90, 146801
(2003).
\bibitem{hasan} M. Z. Hasan and C. L. Kane, Rev. Mod. Phys. 82, 3045
(2010).
\bibitem{kane} C. L. Kane and E. J. Mele, Phys. Rev. Lett. 95,
146802 (2005).
\bibitem{ludwig} A. P. Schnyder, S. Ryu, A. Furusaki and A. W. W. Ludwig, Phys. Rev. B 78, 195125 (2008).
\bibitem{sarma} S. Tewari, T. D. Stanescu, J. D. Sau and S. D.
Sarma, New J. Phys. 13, 065004 (2011).
\bibitem{sau1} J. D. Sau, R. M. Lutchyn, S. Tewari and S. Das Sarma, Phys. Rev. Lett. 104, 040502 (2010).
\bibitem{sau2} S. Tewari, J. D. Sau and S. Das Sarma, Ann. Phys. 325, 219 (2010).
\bibitem{sau3} J. D. Sau, S. Tewari, R. Lutchyn, T. Stanescu and S. Das Sarma, Phys. Rev. B 82, 214509 (2010).
\bibitem{sau4} P. Ghosh, J. D. Sau, S. Tewari and S. Das Sarma, Phys. Rev. B 82, 184525 (2010).
\bibitem{sato} M. Sato and S. Fujimoto, Phys. Rev. B 82, 134521 (2010).
\bibitem{greiner} M. Greiner, O. Mandel, T. Esslinger, T. W.
H\"{a}nsch and I. Bloch, Nature (London) 415, 39 (2002).
\bibitem{moritz} R. J\"{o}rdens, N. Strohmaier, K. G\"{u}nter, H. Moritz and T.
Esslinger, Nature (London) 455, 204 (2008).
\bibitem{lewenstein} M. Lewenstein, A. Sanpera, V. Ahufinger, B. Damski, A. Sen and U.
Sen, Adv. Phys. 56, 243 (2007).
\bibitem{op1} K. Osterloh, M. Baig, L. Santos, P. Zoller and
M. Lewenstein, Phys. Rev. Lett. 95, 010403 (2005).
\bibitem{op2} J. Ruseckas, G. Juzeliunas, P. Ohberg and M. Fleischhauer, Phys. Rev. Lett. 95, 010404 (2005).
\bibitem{op3} X.-J. Liu, M. F. Borunda, X. Liu and J. Sinova, Phys.
Rev. Lett., 102, 046402 (2009).
\bibitem{rf} N. Goldman, I. Satija, P. Nikolic, A. Bermudez, M. A.
Martin-Delgado, M. Lewenstein and I. B. Spielman, Phys. Rev. Lett.,
105, 255302 (2010).
\bibitem{spileman} Y.-J. Lin, K. J.-Garca and I. B. Spielman, Nature
(London) 471, 83 (2011).
\bibitem{k40} J. D. Sau, Rajdeep Sensarma, Stephen Powell, I. B. Spielman and S. Das
Sarma, Phys. Rev. B 83, 140510 (2011).
\bibitem{cross3} H. Hu, L. Jiang, X.-J. Liu and H. Pu, arXiv:1105.2488 (2011).
\bibitem{cross2} Z.-Q. Yu and H. Zhai, arXiv:1105.2250 (2011).
\bibitem{cross1} J. P. Vyasanakere, S. Zhang and V. B. Shenoy, arXiv:1104.5633 (2011).
\bibitem{vbshenoy} J. P. Vyasanakere and V. B. Shenoy,
arXiv:1108.4872 (2011).
\bibitem{melo2} L. Han and C. A. R. S\'{a} de Melo,
arXiv:1106.3613 (2011).
\bibitem{chengang} G. Chen, M. Gong and C. Zhang, arXiv:1107.2627 (2011).
\bibitem{cross4} M. Iskin and A. L. Subas{\i}, arXiv:1106.0473
(2011).
\bibitem{melo} K. Seo, L. Han and C. A. R. S\'{a} de Melo,
arXiv:1108.4068 (2011).
\bibitem{anderson} P. W. Anderson, Science 235, 1196 (1987).
\bibitem{kotliar}  G. Kotliar and A. E. Ruckenstein, Phys. Rev.
Lett. 57, 1362 (1986).
\bibitem{rice} W. F. Brinkman and T. M. Rice, Phys. Rev. B 2,
4302(1970).
\bibitem{kot} A. Georges, G. Kotliar, W. Krauth and M. J. Rozenberg,
Rev. Mod. Phys. 68, 13 (1996).
\bibitem{gutzwiller} M. C. Gutzwiller, Phys. Rev. Lett. 10, 159
(1963); Phys. Rev. 134, A923 (1964); 137, A1762 (1965).
\bibitem{langvin} M. Lavagna, Phys. Rev. B 41, 142 (1990).
\bibitem{iim} A. Imambekov, M. Lukin and E. Demler, Phys. Rev. A 68,
063602 (2004).
\end{thebibliography}
\end{document}